\documentstyle{article}
\def\G{\Gamma}
\def\elvrm{}
\begin{document}
\begin{titlepage}
\centerline{A COMMENT ON FREE-FERMION CONDITIONS}
\centerline{FOR LATTICE MODELS IN TWO AND MORE DIMENSIONS}
\vskip 1truecm
\centerline{J.-M. Maillard$^a$ and C.-M. Viallet$^{a,b}$}
\vskip 1truecm
\centerline{$^a${\em Centre National de la Recherche Scientifique,  
	Universit\'e P. et M.  Curie}}
\centerline{\em L.P.T.H.E. Bo\^{\i}te 126, 4 Place Jussieu, 
	F--75252 Paris Cedex 05}
\vskip .3truecm
\centerline{$^b${\em SISSA, via Beirut 2--4, I--34014 Trieste}}
\vskip 1truecm
\centerline{March 1996}
\vfill
\noindent {\bf Abstract.}
We analyze  free-fermion conditions on vertex models.
We show --by examining examples of vertex models on square, triangular, and 
cubic lattices-- how they amount to degeneration conditions for known
 symmetries of the Boltzmann weights, and propose a general scheme for
 such a process in two and more dimensions.
\vskip 2truecm 
\centerline{{SISSA--Ref. 46/96/EP} \hfil {PAR--LPTHE 96--08} \hfil {hep-th/9603162}}
\vskip .5truecm \hrule\vskip .5truecm
\centerline{e-mail:  maillard@lpthe.jussieu.fr, viallet@lpthe.jussieu.fr}
\end{titlepage}

\section{Introduction}
\null\indent
We consider  vertex models in D dimensions. As is standard, the
bonds of the lattice
carry variables taking $q$ values (colors).
The model is determined by attributing Boltzmann weights to
 the various possible  bond configurations  around  a
 vertex~\cite{LiWu72}.
These homogeneous weights are arranged in a  matrix, which we denote by
$R$.
The size and form of the matrix $R$ vary according to the number of
colors, and the coordination
number of the lattice. Typical examples we will consider are the 2D
square lattice, the 2D
triangular
lattice, and the 3D cubic lattice:
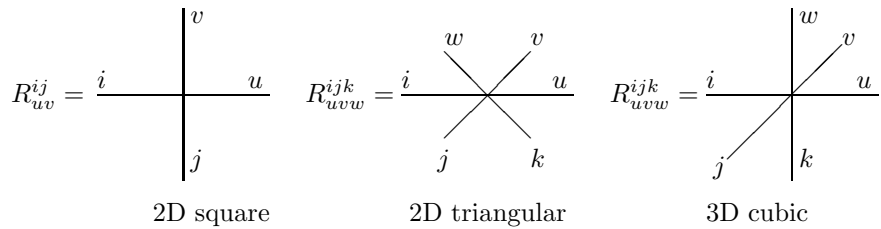
\setlength{\unitlength}{0.0075in}
\begin{eqnarray*}
\begin{picture}(120,120)(20,700)
\thinlines
\put( -40,755){\makebox(0,0)[lb]{\raisebox{0pt}[0pt][0pt]{\elvrm $
R^{ij}_{uv} =   $}}}
\put( 80,820){\line( 0,-1){120}}
\put( 20,760){\line( 1, 0){120}}
\put(125,765){\makebox(0,0)[lb]{\raisebox{0pt}[0pt][0pt]{\elvrm $u$}}}
\put( 85,710){\makebox(0,0)[lb]{\raisebox{0pt}[0pt][0pt]{\elvrm $j$}}}
\put( 85,810){\makebox(0,0)[lb]{\raisebox{0pt}[0pt][0pt]{\elvrm $v$}}}
\put( 20,765){\makebox(0,0)[lb]{\raisebox{0pt}[0pt][0pt]{\elvrm $i$}}}
\end{picture}
\qquad & \qquad
\begin{picture}(120,120)(20,700)
\label{fig1}
\thinlines
\put( 50,730){\line( 1, 1){ 60}}
\put( 110,730){\line( -1,1){60}}
\put( 20,760){\line( 1, 0){120}}
\put(110,795){\makebox(0,0)[lb]{\raisebox{0pt}[0pt][0pt]{\elvrm $v$}}}
\put( 45,710){\makebox(0,0)[lb]{\raisebox{0pt}[0pt][0pt]{\elvrm $j$}}}
\put(125,765){\makebox(0,0)[lb]{\raisebox{0pt}[0pt][0pt]{\elvrm $u$}}}
\put( 50,795){\makebox(0,0)[lb]{\raisebox{0pt}[0pt][0pt]{\elvrm $w$}}}
\put( 110,710){\makebox(0,0)[lb]{\raisebox{0pt}[0pt][0pt]{\elvrm $k$}}}
\put( 20,765){\makebox(0,0)[lb]{\raisebox{0pt}[0pt][0pt]{\elvrm $i$}}}
\put( 15,755){\makebox(0,0)[rb]{\raisebox{0pt}[0pt][0pt]{\elvrm
$R^{ijk}_{uvw}=$}}}
\end{picture}
\qquad & \qquad
\begin{picture}(120,120)(20,700)
\thinlines
\put( 35,715){\line( 1, 1){ 80}}
\put( 80,820){\line( 0,-1){120}}
\put( 20,760){\line( 1, 0){120}}
\put(115,795){\makebox(0,0)[lb]{\raisebox{0pt}[0pt][0pt]{\elvrm $v$}}}
\put( 25,705){\makebox(0,0)[lb]{\raisebox{0pt}[0pt][0pt]{\elvrm $j$}}}
\put(125,765){\makebox(0,0)[lb]{\raisebox{0pt}[0pt][0pt]{\elvrm $u$}}}
\put( 85,710){\makebox(0,0)[lb]{\raisebox{0pt}[0pt][0pt]{\elvrm $k$}}}
\put( 85,810){\makebox(0,0)[lb]{\raisebox{0pt}[0pt][0pt]{\elvrm $w$}}}
\put( 20,765){\makebox(0,0)[lb]{\raisebox{0pt}[0pt][0pt]{\elvrm $i$}}}
\put( 15,755){\makebox(0,0)[rb]{\raisebox{0pt}[0pt][0pt]{\elvrm
$R^{ijk}_{uvw}=$}}}
\end{picture}
\\
\mbox{2D square}\qquad & \qquad \mbox{2D triangular}\qquad & \qquad
\mbox{3D cubic}
\end{eqnarray*}

If the number of colors $q$  is $2$, and we will restrict ourselves to
this case,  then the $R$-matrices are of sizes  $4\times 4$,
$8\times 8$, and $8 \times 8$ respectively. The difference between {2D triangular}
and {3D cubic} for
example does
not show  in the size nor the  form of the matrix.
It will, however, appear in the operations we define on
the matrices.

We shall use a number of elementary transformations acting on the
matrices.
These transformations  come from  the inversion relations and the
geometrical
symmetries of the lattice, in the framework of
integrability~\cite{St79,Ba82,BeMaVi91c,BeMaVi91d},
and beyond integrability~\cite{BeMaVi92}.
They generically form an infinite group
$\G_{lattice}$~\cite{BeMaVi91c,BeMaVi91d}.

The groups  $\G_{lattice}$ have a finite number of involutive
generators.
The first one,  denoted $I$,  is non-linear and does not depend
on the lattice: it is the matrix inversion up to a factor.
The other generators act linearly on $R$, actually by permutations of
the entries, and
represent the geometrical symmetries of the lattice.

For the square lattice, we have two linear transformations, the partial
transpositions $t_l$
and $t_r$~\cite{BeMaVi91d}:
\begin{equation} \nonumber
(t_lR)^{ij}_{uv}= R^{uj}_{iv}, \qquad (t_rR)^{ij}_{uv}= R^{iv}_{uj}
\qquad i,j,u,
v=1..\, q
\end{equation}
The product $t_l t_r$ is the matrix transposition ($l$ stand for `left'
and $r$ stands for `right'
in the standard tensor product structure of $R$).

For the triangular lattice, we have three linear transformations
$\tau_l$, $\tau_m$, $\tau_r$:
\begin{equation}\nonumber
(\tau_lR)^{ijk}_{uvw}= R^{ukj}_{iwv} ,\quad
(\tau_mR)^{ijk}_{uvw}= R^{wvu}_{kji}, \quad
(\tau_rR)^{ijk}_{uvw}= R^{jiw}_{vuk}. \quad
\end{equation}

Finally, for the cubic lattice, we have  three linear transformations
$t_l$, $t_m$,
$t_r$~\cite{BeMaVi91d}:
\begin{equation}\nonumber
(t_lR)^{ijk}_{uvw}= R^{ujk}_{ivw}, \quad
(t_mR)^{ijk}_{uvw}= R^{ivk}_{ujw}, \quad
(t_rR)^{ijk}_{uvw}= R^{ijw}_{uvk}, \quad
\end{equation}
and the product $t_l \; t_m\; t_r$ of the three partial transpositions
is the matrix transposition.

All the generators are involutions.
All products $t_l I$,  $t_m I$ and   $t_r I$  are of infinite order
when acting on a generic matrix, as are $\tau_l I$
and $\tau_r I$.
On the contrary  $\tau_m $ and $I$ commute and $(\tau_m I)^2=1$.

It is straightforward to check that:
\begin{equation}
\label{tegaltau}
t_l \; I \; t_l \;  = \; \tau_l \; I \; \tau_l
\qquad \mbox{and} \qquad
t_r \; I \; t_r \; = \; \tau_r \; I \; \tau_r
\end{equation}
 so that essentially $\G_{triang}$ appears as
a subgroup of $\G_{cubic}$, up to finite factors.

It is important to keep in mind what the ``size'' of the groups $\G$
are. All three $\G_{square}$,
 $\G_{triang}$, and  $\G_{cubic}$ are infinite, but $\G_{square}$ has
 one infinite order generator,
 $\G_{triang}$ has two, and  $\G_{cubic}$ has three.
The last two groups are thus hyperbolic groups~\cite{MaRo94}, and
studying the triangular
lattice can be a good test-case for the more involved tridimensional
cubic lattice.

The groups $\G$  are the building pieces of the group of automorphisms
of
the Yang-Baxter equations and their higher dimensional
generalizations, and solve the so-called 
``baxterization problem''~\cite{BeMaVi91c,BeMaVi91d}.
These equations form  overdetermined systems of multilinear equations,
of which the possible solutions are parametrized by algebraic
varieties~\cite{Ma86}.
The overdetermination increases very rapidly with the dimension of the
lattice.
At the same time, the size of $\G$ also explodes.
When looking at solutions of the Yang-Baxter equations and their
generalizations
to higher dimensional lattices,
one faces a conflict between having a more and more overdetermined
system
and a larger and larger  group of automorphisms for the set of
solutions.
We will show how this conflict  is resolved in some 2D  and known 3D
solutions
by a degeneration of the effective realization of the group $\G$, which
becomes finite.

The content of this letter is the description of
 a mechanism for such a degeneration, obtained by the linearization
of specific elements of $\G$.

We first show how the free-fermion condition on the asymmetric eight
vertex model~\cite{FaWu70}  falls into  this scheme.

We then describe the group $\G_{triang}$  for the 32-vertex model on
the triangular lattice. We show that the free-fermion conditions
given  in~\cite{Hu66,SaWu75} amount to linearizing
the inverse $I$ and make the realization of $\G_{triang}$ finite.

We finally write and discuss similar conditions for the 32-vertex model
on the cubic (3D) lattice, by analyzing solutions of the tetrahedron
equations~\cite{Za81,Ba86,SeMaSt95}.

One of the results we obtain is  that  free-fermion 
conditions should always appear as quadratic conditions, whatever the size 
and form of  the  matrix $R$ is, and in particular whatever the dimension
 and geometry of the lattice are.

There already exists an important literature about free-fermion models.
We may  refer  to~\cite{Sa80,Sa81,Sa81b}, where an
 exploration of the use of grassmannian variables, both for
the construction and the resolution of the models,  can be 
found. 
This work also motivated the interesting  3D construction of~\cite{BaSt84}.

Our approach is based on a direct study of the matrix of Boltzmann weights,
concentrating on the action of the symmetry group $\G$, and provides
another view on this class of  models.

\section{Some notations}
At this point it is useful to introduce some notations  we will use in
the sequel.

We will denote the equality of two matrices $R$ and $R'$ up to an
overall factor by
$R \simeq R'$.

We always denote by $t$ the full matrix transposition.

We will use various gauge transformations (weak graph
dualities)~\cite{GaHi75},
that is to say the conjugation
by invertible matrices which are tensor products, also defined up to
overall factors,
i.e. transformations of the type
\begin{eqnarray*}
R \longrightarrow g_1^{-1}\otimes g_2^{-1} \otimes \dots \otimes
g_k^{-1}
\cdot  R \cdot \;  g_1\otimes g_2 \otimes \dots \otimes g_k \; .
\end{eqnarray*}
Define the  matrices
\begin{eqnarray}
\sigma_0=\pmatrix{ 1 & 0 \cr  0 & 1 }, \quad
\sigma_1=\pmatrix{ 0 & 1 \cr  1 & 0 }, \quad
\sigma_2=\pmatrix{ 0 & -1 \cr  1 & 0 }, \quad
\sigma_3=\pmatrix{ 1 & 0 \cr  0 & -1 }
\end{eqnarray}
and the matrices  $ \sigma_{a_1 a_2 \dots a_k} $
of size $2^k \times 2^k$ ($k$ will be $2$ for the square
lattice, $3$ for the triangular and cubic lattice, and so on) by:
\begin{eqnarray*}
\sigma_{a_1 a_2 \dots a_k} = \sigma_{a_1} \otimes \sigma_{a_2} \otimes
\dots \sigma_{a_k} \; .
\end{eqnarray*}
We denote by $\Sigma_{a_1 a_2 \dots a_k}$ the conjugation by
$\sigma_{a_1 a_2 \dots a_k}$.

Clearly both $t$ and $I$ commute with all  $\Sigma_{a_1 a_2 \dots a_k}$,
up to an irrelevant sign.
Moreover, from the fact that
$
\sigma_a \; \sigma_b \; = \; \pm \; \sigma_b \; \sigma_a, \quad \forall
a,b = 0,1,2,3
$,
the gauge transformations $\Sigma$ satisfy
\begin{eqnarray*}
\Sigma_{a_1 a_2 \dots a_k} \; \Sigma_{b_1 b_2 \dots b_k} \; = \; \pm
\; \Sigma_{b_1 b_2 \dots b_k} \; \Sigma_{a_1 a_2 \dots a_k},
\end{eqnarray*}
meaning that they commute up to a  factor.

Particular gauge transformations of interest are
\begin{eqnarray*}
\pi \; = \; \Sigma_{3 3 \dots 3},
\end{eqnarray*}
and some transformations acting just  by changes of sign of some of the
entries, and denoted
$\epsilon_\alpha$ ($\alpha = l, \; m, \; r,$ ...).

If $k=2$ (square lattice), then $\alpha=l$ or $r$, and
$$ \epsilon_l = \Sigma_{30} , \qquad \epsilon_r = \Sigma_{03}.$$
If $ k=3$ (triangular and cubic lattice):
$$\epsilon_l = \Sigma_{300}, \qquad \epsilon_m = \Sigma_{030}, \qquad
\epsilon_r = \Sigma_{003}. $$

\section{Free-fermion asymmetric eight-vertex model}

The matrix $R$ of the asymmetric eight-vertex model~\cite{Ka74} is of
the form
\begin{equation}
\label{8v}
R = \pmatrix {  a & 0 & 0 & d' \cr
		0 & b & c' & 0 \cr
		0 & c & b' & 0 \cr
		d & 0 & 0 & a' }
\end{equation}
Notice that this form is the most general matrix satisfying $\pi \; R =
R$.

The free-fermion condition~\cite{FaWu70} (see also~\cite{BaSt85abc}) is
\begin{equation}
\label{ff8}
a a' - d d' + b b' - c c' =0
\end{equation}
A matrix of the form (\ref{8v}) may be brought, by similarity
transformations, to a block-diagonal
form
\begin{eqnarray*}
R = \pmatrix{ R_1 & 0 \cr 0 & R_2 }, \qquad\mbox{ with }
\qquad  R_1 = \pmatrix{ a & d' \cr d & a' } \quad \mbox{and}
\quad R_2 = \pmatrix{ b & c' \cr c & b' }.
\end{eqnarray*}
If one denotes by $\delta_1=a a' - d d'$ and  by $\delta_2 =  b b' - c c'$ the
determinants of the two blocks then the matrix inverse $I$  written
polynomially
(namely  $R \rightarrow det(R) \cdot R^{-1}$) just reads
\begin{eqnarray*}
&
a \rightarrow a' \cdot \delta_2, \quad
a' \rightarrow a \cdot \delta_2, \quad
d \rightarrow -d \cdot \delta_2, \quad
d' \rightarrow -d' \cdot \delta_2,& \\
&
b \rightarrow b' \cdot \delta_1, \quad
b' \rightarrow b \cdot \delta_1, \quad
c \rightarrow -c \cdot \delta_1, \quad
c' \rightarrow -c' \cdot \delta_1. \quad &
\end{eqnarray*}
The condition (\ref{ff8}) may be written as  $p_9(R)=0$ with the notations
of~\cite{BeMaVi92}, and is consequently left invariant by $\Gamma_{square}$.
It is straightforward to see that condition (\ref{ff8}) is
$\; \delta_1 = - \delta_2 \;$  and has the effect of linearizing  $I$
into
 \begin{eqnarray*}
&
a \rightarrow a' , \quad
a' \rightarrow a , \quad
d \rightarrow -d , \quad
d' \rightarrow -d' ,& \\
&
b \rightarrow -b', \quad
b' \rightarrow - b , \quad
c \rightarrow c, \quad
c' \rightarrow c'. \quad &
\end{eqnarray*}
The group $\G$ is then realized by permutations of the entries, mixed
with
changes of signs, and its orbits are thus {\em finite}.
The commutators of partial transpositions and inversion, in the sense
of group theory,
i.e: $ t_l  I t_l^{-1} I^{-1} =  ( t_l \; I)^2 $ and
 $ t_r  I t_r^{-1} I^{-1} =  ( t_r \; I)^2 $
 reduce to a change of sign of the non-diagonal entries of $R$.
These commutators are typical infinite order elements of $\Gamma$, when
acting
on a generic matrix, and their degeneration is a key to the finiteness
of the
realization of $\Gamma$.

If we introduce the grading $gr$
\begin{eqnarray*}
gr(R) = \pmatrix { a & 0 & 0 & d' \cr 0 & -b & -c' & 0 \cr
		0 & -c & -b' & 0 \cr d & 0 & 0 & a' }
\end{eqnarray*}
which operates by changing the sign of the entries of only one of the
two blocks,
say $R_2$, then, for any $R$ satisfying (\ref{ff8}), the action of the
inverse reduces to
\begin{eqnarray}
\label{lsq}
I(R) \; \simeq \; t \; \Sigma_{12} \; gr (R)
\end{eqnarray}
where $t$ is  matrix transposition.
In other words we have defined, on all matrices satisfying $\pi (R)
=R$, a linear operator
\begin{eqnarray*}
l_{sq}  = \; t \; \Sigma_{12} \; gr
\end{eqnarray*}
 such that the free-fermion condition (\ref{ff8}) reads
\begin{equation}
I(R) \; \simeq \; l_{sq} (R) \label{lsq1}
\end{equation}
or equivalently
\begin{equation}
R \cdot  l_{sq} (R)  \; \simeq \; \mbox{ unit matrix }
\end{equation}
The linear transformation $l_{sq}$ satisfies a number of relations:
\begin{eqnarray}
l_{sq}^2 = id, \qquad l_{sq} \; t = t \; l_{sq},
\qquad  l_{sq} \; t_\alpha   \; l_{sq} \; t_\alpha = \;
\epsilon_\alpha, \quad \alpha=l,\; r
\end{eqnarray}
Such relations ensure that the orbit of $R$ under $\Gamma$ is finite,
as is readily checked,
and specify the changes of signs to which $(t_l \; I)^2$ and $(t_r \;
I)^2$  reduce.
Notice that the definition of $l_{sq}$ is not unique.

\section{Free-fermion conditions for the 32-vertex model on the triangular
lattice}

We consider the free-fermion conditions for the 32-vertex model on a
triangular lattice, and
 use the notations of~\cite{SaWu75}:
\begin{equation}
\label{rsawu}
R = \left [\begin {array}{cccccccc} {\it f_{0}}&0&0&{\it f_{23}}&0&{\it
f_{13}}&{
\it f_{12}}&0\\0&{\it f_{36}}&{\it f_{26}}&0&{\it f_{16}}&0&0&{\it
{\bar f}_{45}}\\0&{\it
f_{35}}&{\it f_{25}}&0&{\it f_{15}}&0&0&{\it {\bar f}_{46}}\\{\it
f_{56}}&0&0&
{\it {\bar f}_{14}}&0&{
\it {\bar f}_{24}}&{\it {\bar f}_{34}}&0\\0&{\it f_{34}}&{\it
f_{24}}&0&
{\it f_{14}}&0&0&{\it {\bar f}_{56}}
\\{\it f_{46}}&0&0&{\it {\bar f}_{15}}&0&{\it {\bar f}_{25}}&{\it {\bar
f}_{35}}&
0\\{\it f_{45}}&0&0&{\it
{\bar f}_{16}}&0&{\it {\bar f}_{26}}&{\it {\bar f}_{36}}&0\\0&{\it
{\bar f}_{12}}&
{\it {\bar f}_{13}}&0&{\it {\bar f}_{23}}&0&0&{
\it {\bar f}_{0}}\end {array}\right ]
\end{equation}
This matrix may be brought, by a permutations of lines and columns, into
a block diagonal form:
\begin{equation}
\label{blocks} \nonumber
R= \pmatrix{ R_1 & 0 \cr 0 & R_2 }, \qquad \mbox{ with }
\end{equation}
\begin{equation} \nonumber
R_1 = \left [\begin {array}{cccc} {\it f_{0}}&{\it f_{13}}&{\it
f_{12}}&{\it f_{23}}\\{
\it f_{46}}&{\it {\bar f}_{25}}&{\it {\bar f}_{35}}&{\it {\bar
f}_{15}}\\{\it f_{45}}&
{\it {\bar f}_{26}}&{\it {\bar f}_{36}}&
{\it {\bar f}_{16}}\\{\it f_{56}}&{\it {\bar f}_{24}}&{\it {\bar
f}_{34}}&
{\it {\bar f}_{14}}\end {array}\right ]\, ,
 \qquad
R_2 = \left [\begin {array}{cccc} {\it f_{14}}&{\it f_{34}}&{\it
f_{24}}&{\it {\bar f}_{56}}\\{
\it f_{16}}&{\it f_{36}}&{\it f_{26}}&{\it {\bar f}_{45}}\\{\it
f_{15}}&{\it f_{35}}&{\it f_{25}}&
{\it {\bar f}_{46}}\\{\it {\bar f}_{23}}&{\it {\bar f}_{12}}&{\it {\bar
f}_{13}}&{\it {\bar f}_{0}}\end {array}\right ]
\end{equation}
The inverse $I$  written polynomially is  now a transformation of
degree $7$.
If one introduces the two determinants $\Delta_1 = det(R_1) $
and  $\Delta_2 = det(R_2) $, then each term in the expression of $I(R)$
is a product of a degree three minor, taken within a block, times the
 determinant of the other block.

Denoting $\bar{f}_{12}=f_{3456}$ and so on, the free-fermion conditions
of~\cite{Hu66,SaWu75} are:
\begin{eqnarray}
\label{ff32a}
& f_0 f_{ijkl}& =  f_{ij} f_{kl} - f_{ik} f_{jl} + f_{il} f_{jk},
\qquad \forall \; i,j,k,l \, = \, 1,\dots, 6 \\
\label{ff32b}
& f_0 \bar{f}_0 & =  f_{12} \bar{f}_{12} - f_{13} \bar{f}_{13} + f_{14}
\bar{f}_{14} -
f_{15} \bar{f}_{15} + f_{16} \bar{f}_{16}
\end{eqnarray}

What is remarkable is that, not only the rational  variety ${\cal V}$
defined by (\ref{ff32a},\ref{ff32b}), is globally invariant by
$\Gamma_{triang}$, but {\em again} the realization of $\G$ on this
variety is finite.
This comes from
the degeneration of $I$ into a mixture of changes of signs and
permutations of the
entries, as was the case in the previous section.

When relations (\ref{ff32a}, \ref{ff32b}) are satisfied, the action of
$I$ simplifies to
\begin{eqnarray}
\label{ltri}
I(R) \simeq  \; l_{tr} (R)
\end{eqnarray}
with
\begin{eqnarray}
l_{tr} \; = \; t \; \Sigma_{121} \; \simeq  \;  t \; \Sigma_{111} \;
\epsilon_m
\end{eqnarray}

Since we have the prejudice that all free-fermion conditions should
be invariant under  $\G_{triang}$, as  (\ref{ff32a}, \ref{ff32b}) are,
one should complement (\ref{ltri}) with
\begin{eqnarray}
I \; \tau_l (R) & \simeq  &  l_{tr}\;  \tau_l \; (R)   \label{ltril}
\\
 I \; \tau_r (R) & \simeq  &  l_{tr} \; \tau_r \; (R) \label{ltrir}
\end{eqnarray}

We may list some useful relations:
\begin{eqnarray}
\label{sioux00}
& \tau_l \; \Sigma_{abc} = \Sigma_{acb} \; \tau_l , \qquad
\tau_m \; \Sigma_{abc} = \Sigma_{cba} \; \tau_m , \qquad
\tau_r \; \Sigma_{abc} = \Sigma_{bac} \; \tau_r  \qquad  &\\
\label{sioux01}
& t \; \tau_\alpha = \tau_\alpha \; t  \qquad  \qquad
\forall \; a,b,c = 0,1,2,3, \; \forall \; \alpha=l,m,r & \\
\label{sioux1}
  & ( \tau_\alpha \; \tau_\beta \; \tau_\gamma)^2 = id \qquad\forall
  \alpha, \beta, \gamma = l,m,r  & \\
\label{sioux2}
& \tau_l \tau_m \tau_r \; I  =  I \;  \tau_l \tau_m \tau_r,
\qquad\qquad
\tau_l \tau_r \tau_l \; I  =  I \; \tau_l \tau_r \tau_l  &
\end{eqnarray}
The linear transformation $l_{tr}$ satisfies in addition:
\begin{eqnarray}
& l_{tr}^2=id, \qquad l_{tr} \; t = t \; l_{tr} & \label{rell1} \\
& l_{tr}  \; t_l \; l_{tr} t_l = \epsilon_l, \qquad \label{rell2}
l_{tr}  \; t_m \; l_{tr} t_m = id , \qquad
l_{tr}  \; t_r \; l_{tr} t_r = \epsilon_r &
\end{eqnarray}

Using relations (\ref{sioux00}) to (\ref{rell2}), it is possible to
show  that
the completed system  (\ref{ltri},~\ref{ltril},~\ref{ltrir})
is  left invariant by the action of the group $\G_{triang}$.

The system is also invariant  under  the gauge transformations
leaving the form (\ref{rsawu}) stable.
(Hint: The gauge transformations
leaving the form (\ref{rsawu}) stable satisfy
$ \; {}^t\gamma \; \Sigma_{abc} \; \gamma \; \simeq \; \Sigma_{abc} \;
$
when $a,b=1,2$).

Moreover  the  generic 32-vertex  invertible solutions of  the
completed system (\ref{ltri},~\ref{ltril},~\ref{ltrir}) satisfy the
free-fermion conditions  (\ref{ff32a},~\ref{ff32b}).

It is clear from  (\ref{ltri},~\ref{ltril},~\ref{ltrir})
and    (\ref{sioux00}) to (\ref{rell2}) that the realization of the
group $\G_{
triang}$
is finite, when conditions (\ref{ff32a}, \ref{ff32b}) are fulfilled.

One should also notice that any linearization condition of the type of
(\ref{ltri})
 is a set of {\it quadratic conditions, whatever the size of the matrix
 is}.
Indeed they mean that the matrix product of $R$ with  some linear
transformed
${\cal L}(R)$ of $R$
is proportional to the unit matrix, i.e:
\begin{eqnarray}
R \cdot {\cal L}(R) \simeq \; \mbox{unit matrix}
\end{eqnarray}
 and this is a set of  quadratic conditions.

{\bf Remark}: The invertible solutions of (\ref{ltri}) form a group for
the ordinary
matrix product, since $I \cdot l_{tr}$ is an automorphism of the group of
invertible
matrices of the form (\ref{rsawu}), i.e.
$I(l_{tr} (R_1 \cdot R_2)) =I( l_{tr}(R_1)) \cdot I( l_{tr}(R_2))$.
The extra conditions added when completing
the system break this in such a way that the ordinary matrix product of
three solutions is
another solution.
In other words, if $R_1, \; R_2, \; R_3 \; \in {\cal V}$, then
$ R_1 \cdot R_2 \cdot R_3 \; \in {\cal V} $, while $R_1 \cdot R_2 \notin {\cal V}$.
 This was actually already the case for solutions of (\ref{ff8}), but
 the mechanism is more subtle here
as conditions  (\ref{ff32a}, \ref{ff32b}) imply $\Delta_1=+ \Delta_2$.

\section{32-vertex model on the cubic lattice}

We now turn to a solution of the tetrahedron
equations~\cite{SeMaSt95,Ko94,Hi94}.
Let $R$ be of the form
\begin{eqnarray}
\label{serg}
R = \left (\begin {array}{cccccccc}
d&0&0&-a&0&-b&c&0\\0&w&x&0&y&0&0&z\\0&
x&w&0&z&0&0&y\\-a&0&0&d&0&c&-b&0\\0&-y&z&0&w&0&0&-x\\b&0&0&c&0&d&a&0\\
c&0&0&b&0&a&d&0\\0&z&-y&0&-x&0&0&w\end {array}\right )
\end{eqnarray}
The form of  (\ref{serg}) is stable \footnote {Notice that the form
(\ref{serg}) is not stable by the circular permutation of the three spaces
$\{l,m,r\}$.}  under the group $\Gamma_{cubic}$, and it is natural to look
for invariants of $\Gamma$ in the space of parameters
$\{a,b,c,d,x,y,z,w\}$~\cite{FaVi93}.
There exist five algebraically independent quadratic polynomials in the
entries, transforming
covariantly, and with the same covariance factors under all generators
of $\Gamma_{cubic}$.
They are:
$$ ax, \quad by, \quad cz, \quad dw ,\quad \mbox{ and } \quad
Q = a^{2}+c^{2}-d^{2}-y^{2}-b^{2}+x^{2}-w^{2}+z^{2}\; . $$
We thus have four algebraically independent invariants of
$\Gamma_{cubic}$, say for example
$$ \chi_1 = {{ax}\over{dw}}, \quad
\chi_2 = {{by}\over{dw}}, \quad
\chi_3 = {{cz}\over{dw}}, \quad \mbox{and} \quad
\chi_0 = {{Q}\over{dw}}. $$
A complete analysis   shows that there is no other algebraically
independent invariant of
$\Gamma_{cubic}$. A  numerical and graphical
study~\cite{BeMaVi91e}, shows
 how ``big'' the realization of $\Gamma$ is for generic values of the
 above invariants.

These invariants are completely specified in the
solution~\cite{SeMaSt95}, for which
\begin{eqnarray}
& \chi_1 \; = \; \chi_2 \; =  \; \chi_3 \; = \; 1  & \label{chi1} \\
&  \chi_0 =0 & \label{chi0}
\end{eqnarray}
Out of the four invariants, $\chi_0$ plays a special role.
 {\em If $\chi_0=0$, then the
action of $I$ linearizes quite in the same way as in the previous
cases}.
Notice that, strictly speaking, condition (\ref{chi0}) is not so much
an assignment of value to the invariant $\chi_0$ but rather a
 vanishing condition for  the covariant quantity $Q$. 
Recall that assigning a definite value to an 
invariant  object is meaningful whatever this value is. 
On the contrary covariant objects cannot
be assigned a value unless this value is zero.

When $Q=0$, one gets
\begin{eqnarray}
\label{lcub}
I(R) \; \simeq \; l_{c}(R) 
\end{eqnarray}
The linear transformation $l_{c}$ may be written
\begin{eqnarray*}
 l_{c} = \; t \; \Sigma_{030} \; gr,
\end{eqnarray*}
  where $t$ is  transposition and $gr$ is
a grading changing the sign of the entries of $R$ belonging to the same
block, say
$\{x,y,z,w\}$.
Notice that the definition of $l_{c}$ is not unique, due to the very
specific form of
(\ref{serg}).
Notice also that $Q =0$ is one of  two quadratic conditions
ensuring the equality of the
determinants of the two blocks of $R$ (see~(\ref{blocks})). The other one
 is not stable under $\Gamma$.

The linear transformation $l_{c}$ satisfies
\begin{eqnarray}
& l_{c}^2=id, \qquad l_c \; t = t \; l_{c}, & \label{relc1} \\
& l_{c}  \; t_\alpha \; l_{c} t_\alpha = \epsilon_\alpha, \qquad
\forall \alpha=l,\; m, \; r &
\label{relc2}
\end{eqnarray}

Any matrix of the form (\ref{serg}) with $Q=0$ obeys
\begin{eqnarray}
& I(R) \; \simeq \; l_{c} (R) & \label{lintet1} \\
& I \; t_\alpha \; (R) \; \simeq \; l_{c}\; t_\alpha \; (R) \qquad
\forall \alpha = l,m,r &
\label{lintet2}
\end{eqnarray}
Using (\ref{relc1},\ref{relc2}), it is straightforward to show that
the  complete
system (\ref{lintet1},\ref{lintet2}) is invariant under $\Gamma_{cubic}$
and
that the  orbit of $R$ is finite.\footnote{This is also the case for 
the bidiagonal  solution of the ``constant'' tetrahedron equations 
of~\cite{Hi93b}.}

The study of the additional conditions (\ref{chi1}) would take us
beyond the scope of this letter,
but we may make a few remarks.

The first remark is that since among  conditions  (\ref{chi1},\ref{chi0}),
only  (\ref{chi0}) has to do with the finiteness of the realization
of $\G$,  (\ref{chi1}) may have nothing to do with
 free-fermion conditions.
They  are {\em additional constraints} making the resolution
of the tetrahedron equations possible, and this may be understood as follows.

The tetrahedron equations are in essence a compatibility condition for
the existence of
non-trivial solutions of the  ``propagation properties''~\cite{Ba73}
(alias ``Zamolodchikov
algebra''~\cite{ZaZa79}, alias ``vacuum curves''~\cite{Kr81,Ko94},
alias ``pre-Bethe Ansatz''
equations~\cite{BeMaVi92,BeBoMaVi93}):
\begin{equation}
\label{pba}
R \; \pmatrix{1 \cr p} \otimes \pmatrix{1 \cr q} \otimes \pmatrix{1 \cr
r} \; \simeq \;
\pmatrix{1 \cr p'} \otimes \pmatrix{1 \cr q'} \otimes \pmatrix{1 \cr
r'}
\end{equation}
What conditions (\ref{chi1},\ref{chi0})  ensure is the existence, for
fixed $R$,
 of a {\em one-parameter family of solutions} of~(\ref{pba}).
In the case we consider here, the family happens to be parametrized by a 
{\em  curve of  genus larger than one}.

By eliminating  $\{q,q',r,r'\}$ (resp.  $\{p,p',r,r'\}$ or
$\{p,p',q,q'\}$)
from (\ref{pba}), one gets conditions relating $p,p'$, (resp. $q,q'$
and $r,r'$).
Such  relations are generically of degree 8 (biquartics).
One effect of  (\ref{chi1},\ref{chi0})  is that they all reduce to 
{\em asymmetric}  biquadratic relations, defining three genus one  curves 
of the form
\begin{eqnarray} \label{asym}
&  x^2 y^2 -1 +  ( y^2 - x^2) \; \kappa_{xy} =0  & 
\end{eqnarray}
\begin{eqnarray*}
 & \mbox{with} \quad \kappa_{pp'}= {\displaystyle {b c \; ( d^2-a^2)}
\over{\displaystyle a d \; (b^2-c^2)}} , \quad
\kappa_{qq'} = - {\displaystyle {a c \; (b^2+d^2)}
\over{\displaystyle  b d \; (a^2 + c^2)}}, \quad
\kappa_{rr'} = {\displaystyle  { a b\;  (c^2-d^2)}
\over{\displaystyle  c d \; (a^2-b^2) }}. &
\end{eqnarray*}
These three elliptic curves have different (algebraically independent)
moduli.
Their asymmetric character may be an obstacle to the use of (\ref{pba})
in the construction of the Bethe Ansatz states~\footnote{R.J. Baxter,
private communication.}, since the composition of relations of type 
(\ref{asym})
reproduces the same type of relations, but alters the value of $\kappa$ by:
\begin{eqnarray} \label{land}
\kappa \longrightarrow {1\over 2} \left( \kappa + {1\over{\kappa}} \right)
\end{eqnarray}
Exceptional values of $\kappa \; (\pm 1, \infty)$, yielding a rationalization 
of (\ref{asym}), are fixed points of (\ref{land}). 
For these exceptional values, in particular $\kappa=\infty$,  
obtained with $d=0$, 
the construction of a 3D Bethe Ansatz may be  envisaged.

\section{Conclusion}
We have  shown, through specific examples,  how free--fermion conditions 
turn into degeneration conditions of our groups $\G$: 
the generically  non-linear (rational) infinite  realization of $\Gamma$ 
becomes a linear  finite group.

We believe this is a characteristic feature of 2D free-fermion models. 

We have shown that the known vertex solution of the tetrahedron equations 
does  have such a feature.
An appealing issue is to decide whether or not such a statement can be 
made about other 3D  and higher dimensional  models.
Of course the full answer will come from linking directly the phenomenon
we describe with  explicit calculus  using grassmannian variables.
The particularly simple form of the conditions (combinations of
products of entries with plus and minus signs), 
and the  linearization  process of the inverse
should stem from elementary properties of exponentials of quadratic forms
in anticommuting variables.

Producing new solutions of the tetrahedron equations is another  
challenging problem.
What could be done is to look for forms of the matrices $R$ enjoying the 
linearization property we have described. This is a rather simple 
way to produce ``reasonable'' Ans\"atze for $R$.

The next step would then be to study the so-called propagation properties
(see above) rather than confronting directly  the tetrahedron equations
themselves. 

Indeed these simpler equations, because they govern the construction of 
Bethe Ansatz  states --a basic in the field--,   underpin 2D, 3D, and higher
dimensional integrability.

\bigskip
{\bf Acknowledgments}: {JMM would like to thank Pr. R.J. Baxter for an
invitation at the  Mathematics Dept. A.N.U., Canberra, and for many fruitful 
discussions. JMM would also like to thank
Pr. F.Y.Wu for interesting discussions on the 32-vertex model.
CMV would like to thank the members of the High Energy Physics group of SISSA, 
 Trieste,  where this work was completed, for their warm hospitality, and
 acknowledges support from the EEC program ``Human Capital and Mobility''.
Both authors benefited from stimulating discussions with M. Bellon and M.
Talon.}

\end{document}